\documentclass[twocolumn,secnumarabic,amssymb,amsmath,nofootinbib,tightenlines,nobibnotes,aps,prl,showpacs]{revtex4}
\usepackage{graphicx,epsfig}

\newcommand{\lsi}{\raise0.3ex\hbox{$<$\kern-0.75em\raise-1.1ex\hbox{$\sim$}}}
\newcommand{\gsi}{\raise0.3ex\hbox{$>$\kern-0.75em\raise-1.1ex\hbox{$\sim$}}}
\begin{document}

\def\gsi{$^{1}$}
\def\hdpi{$^{2}$}
\def\hdth{$^{3}$}

\title{Chemical Freeze-out and the QCD Phase Transition Temperature}

\author{P. Braun-Munzinger\gsi, J. Stachel\hdpi, Christof Wetterich\hdth}

\affiliation{
\gsi GSI Darmstadt, Germany\\
\hdpi Physikalisches Institut, Universit{\"a}t Heidelberg, Germany\\
\hdth Institut f{\"u}r Theoretische Physik, Universit{\"a}t Heidelberg, Germany}

\begin{abstract}
We argue that hadron multiplicities in central high energy
nucleus-nucleus collisions are established very close to the phase
boundary between hadronic and quark matter. In the hadronic picture
this can be described by multi-particle collisions whose importance is
strongly enhanced due to the high particle density in the phase
transition region.  As a consequence of the rapid fall-off of the
multi-particle scattering rates the experimentally determined chemical
freeze-out temperature is a good measure of the phase transition
temperature.

\end{abstract}


%\hfill{HD-THEP-03-55}

\maketitle

%%%%%%%%%%%%%%%%%%%%%%%%%%%%%%%%%%%%%%%%%%%%%%%%%%%%%%%%%%%%%%%%%%%%%%%%

The yield of (multi-) strange hadrons produced in central high energy
nucleus-nucleus collisions was proposed two decades ago \cite{raf1} as
a signature of quark-gluon plasma (QGP) formation. Hadron yields observed in
such collisions at AGS, SPS, and RHIC energies are found to be
described with high precision within a hadro-chemical equilibrium
approach \cite{agssps,satz,heppe,cley,beca1,rhic,nu,beca2,rapp}, governed
by a chemical freeze-out temperature T$_{ch}$, baryo-chemical
potential $\mu$ and the fireball volume V$_{ch}$. A recent review can
be found in \cite{review}.  Importantly, the data at SPS and RHIC
energy comprise multi-strange hadrons including the $\Omega$ and $\bar
\Omega$. Their yields agree with the chemical equilibrium calculation
and are strongly enhanced as compared to observations in pp
collisions. The time needed to achieve this equilibrium for $\Omega$
baryons via two-body collisions was estimated \cite{raf1} to be much
longer than reasonable lifetimes of the fireball. The observations
were thus interpreted as a sign that the system had reached a partonic
phase prior to hadron production \cite{gerry,stock,heinz1}.

In this note we argue that the chemical freeze-out temperature
T$_{ch}$ is actually very close to the critical temperature T$_c$ of
the QCD phase transition. This observation has a far reaching
consequence: Since T$_{ch}$ has been measured for different values of
$\mu$ the approximate association of T$_c$ with T$_{ch}$ implies that
we have experimental knowledge of part of the critical line in the
QCD-phase diagram.

Let us first sketch our overall picture and detail our arguments
subsequently.  Hadro-chemical equilibration is achieved during or at
the end of the phase transition. In particular, the number of strange
quarks may be established in the plasma phase and/or hadronization of
the QGP. During the very early stages of the hadronic phase the
relative numbers of strange baryons and mesons $K,K^*,\Lambda, \Sigma,
\Xi, \Omega$ are then realized at the thermal equilibrium values according
to the Bose-Einstein or Fermi distribution
\begin{equation}\label{eq1}
n_j=\frac{g_j}{2\pi^2}\int^{\infty}_0 p^2dp\{\exp[(E_j(p)-\mu_j)/\rm{T}]\pm1\}^{-1}
\end{equation}
Here $E^2_j=M^2_j+p^2$ and $M_j$ proportional to the vacuum mass of
the hadron $j$, $\mu_j$ is the effective chemical potential, and $g_j$
counts degrees of freedom (for details see \cite{review}). The high
accuracy of the distribution (\ref{eq1}) in reproducing the data
suggests that $T_{ch}$ plays effectively the role of a universal
temperature, governing simultaneously the chemical and kinetic
distributions.

In the hadronic picture the production of multi-strange hadrons can be
described by multi-hadron strangeness exchange
reactions\footnote{Production of multi-strange baryons by
multi-particle collisions has also been considered by
\cite{cgreiner}. Their argument focuses on anti-hyperon production at
high baryon density, where indeed relatively short equilibration times
are obtained. The authors conclude that their approach should not be
applicable for RHIC energies, unless the hadronic phase has a rather
long lifetime. Furthermore this approach does not take account of the
expected rapid change of density near a phase transition which is
central for our argument.}. The multi-hadron scattering is
substantial, however, only in the immediate vicinity of the critical
temperature T$_c$. As T decreases the multi-particle rates drop very
rapidly with a high power of the particle density. Below some
temperature T$_{ch}$ very close to T$_c$ only two-particle
interactions and decays remain as relevant processes for a change in
the relative particle numbers. These are too slow in order to
equilibrate the distributions or to catch up with the decreasing
temperature - chemical freeze-out occurs for T$_{ch}\approx$ T$_c$.
We proceed to discuss the three main points of this scenario in the
following in more detail.

\begin{figure}[hbt]
\begin{center}
\vspace{-1.0cm}
\includegraphics[bbllx=120,bblly=10,bburx=400,bbury=520,width=5.0cm]{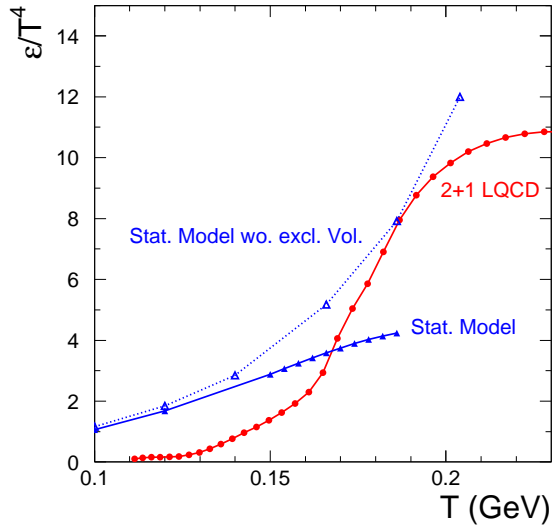}
\vspace{-.80cm}
\caption{Comparison of the temperature dependence of energy density  in lattice calculations
  with 2 light and 1 heavier quark flavor \cite{karsch} with results from the
  hadronic gas model of \cite{agssps,heppe} with and without repulsive interactions.}
\label{lattice}
\end{center}
\end{figure}

\medskip
(A) The QCD phase transition corresponds to a change in the effective
degrees of freedom (from hadrons to quarks and gluons) in a narrow
temperature interval \footnote{ Our central argument will make no
distinction between a true phase transition and a rapid
''crossover''.}. For both the hadronic and quark-gluon phases
sufficiently far away from T$_c$ the dominant processes in thermal
equilibrium are two-particle scattering and decays. This is consistent
with an effective (pseudo-)particle description.  Close to T$_c$,
however, collective phenomena play an important role. (As an example,
near T$_c$ the $\sigma$-resonance may behave like a particle with mass
almost degenerate with the pions.) If $T_{ch}$ is close to $T_c$ the
chemical equilibration may be described in several equivalent
pictures: multi-hadron scattering, time evolving classical fields or
hadronization.  We emphasize, however, that no picture should
contradict a hadronic description.  In the end, the chemical
equilibrium distribution has to be established by fast processes
involving hadrons (not quarks and gluons).

Let us approach the phase transition (or a rapid crossover) from the
hadronic phase. For T sufficiently below T$_c$ not much happens on the
level of microscopic scattering processes between hadrons. The main
effect of an increase of the temperature is an increase of the
density. Near T$_c$, however, the density is so high that new dynamics
can be associated with collective excitations. It is precisely the
behavior of these collective excitations that triggers the
transition. On the level of individual hadrons the propagation and
scattering of collective excitations is expressed in the form of
multi-hadron scattering.
Since the collective dynamics becomes dominant only near
T$_c$ the same holds for the multi-hadron scattering.

We next argue that the temperature range where multi-hadron processes can 
dominate is actually very narrow (typically a few MeV). This is due to
(i) a rapid increase of the particle densities as a function of T
and (ii) a very steep dependence of multi-particle scattering rates on
the density of the incoming particles. Evidence for rapid energy
density changes near T$_c$ comes from recent lattice QCD results
\cite{karsch}. This is shown in Fig.~\ref{lattice}, where the observed
rapid rise in energy density beyond the simple T$^4$ dependence
reflects the large increase in degrees of freedom. The statistical
model of a hadron resonance gas \cite{agssps,heppe} exhibits only a
moderate increase beyond the T$^4$ behavior (see Fig.~\ref{lattice}),
determined by an interplay between the relevant number of hadronic
states and the increased importance of repulsive interactions modeled
by an excluded volume correction\footnote{Neglecting the repulsive
interactions the energy density of a hadronic gas diverges, reflecting
the Hagedorn temperature (see dotted line in Fig.~\ref{lattice} and
\cite{karsch1})}.
% The lattice calculations still employ
%unrealistically large quark masses, leading to much too small energy
%densities at low temperatures, compared to the hadronic gas
%values. The relevant feature for our current argument is, however, the
%rapid increase near T$_c$, implying that there also the particle
%density will change very rapidly.

\begin{figure}[!]
\begin{center}
\vspace{-3.20cm}
\includegraphics[bbllx=20,bblly=7,bburx=350,bbury=450,width=9.0cm]{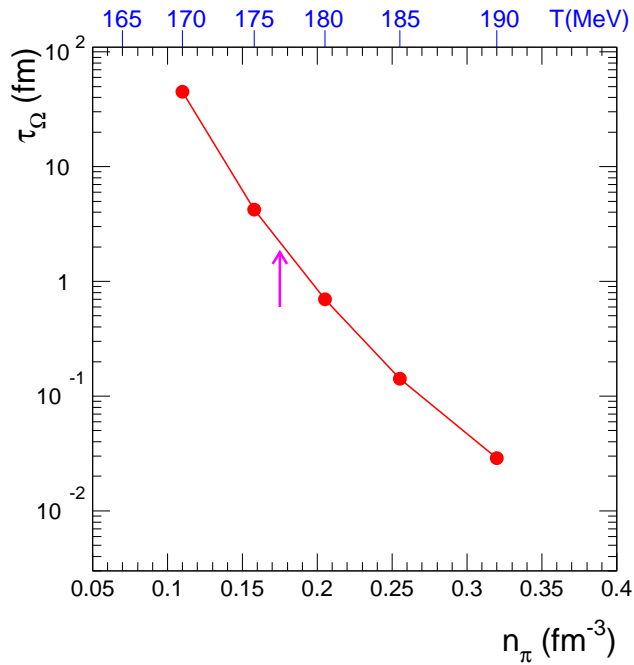}
\vspace{.20cm}
\caption{Time $\tau_{\Omega}=n_\Omega/r_\Omega$ needed to bring $\Omega$ baryons into
chemical equilibrium via multi-particle collisions. We display the
dependence on the pion density $n_\pi$ and on T. The temperature scale
is obtained as discussed in the text. The arrow indicates the chemical
freeze-out temperature T$_{ch}=176$ MeV.}
\label{tau}
\end{center}
\end{figure}

The rate of an individual multi-particle scattering process with
$n_{in}$ incoming particles and average density $\bar{n}$ grows
as $\bar{n}({\rm T})^{n_{in}}$. In addition, also the phase space
increases with temperature. In the temperature region of
multi-particle scattering dominance one expects that many processes
(with different $n_{in}$) become of similar strength, thus increasing
again the total rate. For the purpose of demonstration we model the
temperature dependence of a typical rate (say for scatterings which
change the number of $\Omega$-baryons) by $r\sim \bar{n}({\rm
T})^\gamma$, with $\gamma \gg 1$. Within a narrow temperature interval
$\Delta T=5$~MeV this rate changes by a substantial factor
$(1+\frac{\Delta {\rm T}}{\rm T})^{\beta\gamma}$ with
$\beta=d\ln\bar{n}/d\ln {\rm T}\gg 1$. This is demonstrated in Fig.~\ref{tau}.
At T$_c$, the multi-particle scattering rates are substantial. For
${\rm T<T}_c$, however, they drop so rapidly that only a small
temperature interval around T$_c$ is left where the multi-particle
scattering processes can dominate. The evaluation of these
multi-particle rates is detailed in point (C) below.

\medskip

(B) Let us next turn to the issue of chemical equilibrium and argue
that two-particle scattering is insufficient to achieve or sustain
it. We focus in the following on the analysis of data at RHIC energies
and comment at the end on the applicability of our considerations at
lower energies. The high accuracy of the statistical model predictions
in reproducing experimental particle ratios varying over several
orders of magnitude \cite{rhic,review} should be taken as an indication that
hadro-chemical equilibration has occurred in the system. Further,
chemical equilibration according to eq.~(\ref{eq1}) requires two crucial
ingredients: (i) The particle number changing reaction rates must be
sufficiently high such that the particle distributions can adapt to
a given $T$. 
(ii) At freeze-out the masses of the different hadrons must be
proportional to their vacuum masses. From (ii) we conclude immediately
that chemical equilibration and freeze-out occur in the hadronic
phase, T$_{ch} \leq$ T$_c$. The particle distribution in the
quark-gluon plasma has no memory of the individual hadronic vacuum
masses - for example, the relative density of strange particles (at
the same given $\mu$ and T) would be determined by the strange quark
mass m$_s$ rather than by the individual masses of
$K,K^*,\Lambda,\Sigma,\Xi,\Omega$. 
			
To demonstrate that T$_{ch}$ is close to T$_c$ we exclude two possible
alternative scenarios with T$_{ch}$ substantially smaller than T$_c$:
First, we show that an extended period of ``hadronic chemical
equilibrium'' with T$_{ch}<$T$<$T$_c$ is quite unlikely. Second, we
generalize this argument to demonstrate that ``late equilibration'' at
a temperature close to T$_{ch}$ but significantly smaller than T$_c$
is also disfavored.

The first argument is conceptually simple since the condition for
chemical equilibrium at a temperature T substantially smaller than
T$_c$ can be formulated by using the equilibrium rates and
distributions. For our purpose complete thermalization of all
quantities is not necessary - the ''prethermalization'' \cite{preth}
of some rough quantities like relative particle densities and
approximate momentum distributions is sufficient,such that it makes
sense to speak about temperature and chemical potential
(in the sense of eq. (\ref{eq1})) and to compute rates in thermal
equilibrium.

We first need the relevant time scale to which the two-particle
scattering rates have to be compared. In equilibrium this is given by
the inverse rate of decrease of temperature $\tau_T$.
For this purpose we assume, guided by
recent results from two-pion correlation measurements
\cite{ceres2pi1,ceres2pi2,rhic2pi,hbtreview}, two possible scenarios
for the evolution of the fireball. In both cases we use the
observation that the density decreases by only 30\% between chemical
and thermal freeze-out and our knowledge of T$_{ch}$ = 176 MeV. From
the two-pion correlation data we obtain, for a central rapidity slice,
the transverse and longitudinal radii at thermal freeze-out of 5.75
and 7.0 fm, a longitudinal expansion velocity $\beta_{\parallel}$ = 1,
and a transverse expansion velocity $\beta_{\perp}$=0.5. The thermal
freeze-out radii are to be understood as widths of Gaussian
distributions and give a volume of V$_f$= 3650 fm$^3$. Scenario (1)
assumes that the shape of the density distributions is the same
(i.e. Gaussian) at thermal and chemical freeze-out and that
accordingly an increase in density by 30 \% is equivalent to a 30 \%
decrease in volume corresponding to V$_{ch}$=2600 fm$^3$. In scenario
(2) we use an initial volume of 1450 fm$^3$ corresponding to a flat
distribution over one unit of rapidity. With the assumption of
isentropic expansion at the above velocities we find the longitudinal
and transverse radius parameters at T$_{ch}$ as well as the duration
of the expansion and the thermal freeze-out temperature T$_f$. For
scenarios (1) and (2) the duration of the expansion in the hadronic
phase is $\tau_f$ = 0.9 and 2.3 fm and the thermal freeze-out
temperature is T$_f$ = 158 and 132 MeV. Consequently, in the hadronic
phase near T$_{ch}$ the rate of decrease in temperature may be
estimated as $|\dot{\rm{T}}/\rm{T}|=\tau^{-1}_T=(13 \pm
1)$\%/fm$=(7.7\pm0.6\, {\rm fm})^{-1}$.  Note that these time scales
are entirely consistent with the duration of pion emission, also
obtained from the two-pion correlation data, of less than 2 fm. In
both scenarios the lifetime of the fireball is rather short, leaving
little room for an extended period of ``hadronic cooking'' at
temperatures significantly below T$_{c}$.

As a simple example, a decrease in temperature by $\Delta$~T=5 MeV
reduces the equilibrium ratio of $\Omega$-baryons over kaons,
$n_{\Omega}/n_K$, by a factor $F_{\Omega K}=$ 1.13 (using thermal
model densities of \cite{rhic}). Adaptation of the particle
distribution to the changing temperature requires $F_{\Omega
K}=\exp(\alpha\Delta t)=\exp(\alpha\tau_{\rm T}\Delta \rm{T/T}_{ch})$
with $\alpha=d\ln(n_{\Omega}/n_K)/dt$. Let us define the rate of
change of individual particle number densities as
\begin{equation}
\bar{r}_j=\frac{\dot{N}_j}{V}=\dot{n}_j+n_j\dot{V}/V.
\end{equation}
(The last term accounts for the decrease in particle number densities due to the
 volume change.) Maintaining chemical equilibrium needs
\begin{equation}\label{3}
\left|\frac{\bar{r}_{\Omega}}{n_{\Omega}}-\frac{\bar{r}_K}{n_K}\right|=
\frac{\ln F_{\Omega K}}{\tau_T}\frac{\rm{T}_{ch}}{\Delta
\rm{T}}=(1.10-0.55)/{\rm fm}=0.55/{\rm fm}.
\end{equation}
The numerical evaluation of the two terms in the difference agrees well
with the direct evaluation of the term involving $F_{\Omega K}$ in
eq.~(\ref{3}). For the difference in the rate of relative density change of
$\Omega$-baryons to protons we obtain similarly a value of
(1.10-0.90)/fm=0.20/fm. Both rates are evaluated here at
T$_{ch}$. Eq. (\ref{3}) can be obeyed either by the destruction of
$\Omega$-baryons or the production of kaons. Since $\Omega$-baryons
decay weakly the effect of decays is completely negligible.  A typical
two-particle scattering $\Omega+\bar{K}\to\Xi+\pi$ (or
$\Omega+\pi\to\Xi +K$) yields at most
$\bar{r}_{\Omega}/n_{\Omega}=n_{\bar K}\langle v \sigma
\rangle$=0.018/fm. For this estimate we used a (large) strangeness
exchange cross section of $\sigma$=10 mb and a relative velocity of
$v$=0.6. Similarly, for a typical kaon production process
$\pi+\pi\to K+\bar{K}$ one would obtain
$\bar{r}_K/n_K=0.18/$fm, using the measured cross section of 3 mb
\cite{strange}. Clearly both numbers are much too small to maintain
equilibrium close to T$_{ch}$. Analogous arguments can be made for the
$\Omega$/p ratio. The reaction $\pi^+ + \Sigma^- \to K^- + p$
contributes to the rate of relative density change of protons a value
of 0.018/fm. One would need about 50 reactions with similar cross
section (12 mb, \cite{strange1}) to keep the proton density in
equilibrium. For the $\Omega$ baryons there is obviously no way to
achieve this.  We conclude that two-particle reactions and decays are
not fast enough to maintain chemical equilibrium for multi-strange
baryons in the hadronic phase near T$_{ch}$. This finding is also
supported by studies using cascade codes \cite{cgreiner1} and rate equations
\cite{kapusta}.

We next turn to the production rates for situations where thermal
equilibrium has not yet been realized, e.g. during hadronization. The
production rates should be consistent between different pictures for
such processes and permit a hadronic description, at least towards the
end of the chemical equilibration process. Therefore our picture does
not require a detailed understanding of hadronization. Although our
estimates of $\bar{r}_j/n_j$ have used thermal distributions for the
incoming particles our arguments can be extended to non-thermal
situations: there is no reason why the rates should be much larger or
the available times longer.  In the final approach to chemical
equilibrium (needed in order to achieve the high accuracy of the
observed thermal description of the data) the densities of incoming
particles must already be close to equilibrium. Also only very rough
features of the momentum distribution of the incoming particles are
needed for an estimate of the magnitude of $\bar{r}_j/n_j$.
Therefore, the two-particle scattering is also too slow to {\it
achieve} chemical equilibration in the production of multistrange
hadrons. In particular, this applies to a possible picture of chemical
equilibration by hadronization: hadronization at T much below T$_c$
would not lead to equilibrium abundances since there is not enough
time to produce the multistrange baryons with rates corresponding to
T.( Otherwise the hadronization picture would be in clear discrepancy
with an equivalent hadronic picture for which the rates are dominated
by two particle scattering.)  This closes our argument: At T$_{ch}$
either multi-particle interactions must be important or the cross
section must be dramatically larger than in the vacuum. Both
possibilities are conceivable only for T$_{ch}$ very close to T$_c$.

\medskip

(C) The chemical equilibration of hadrons should be accessible to a
hadronic description, at least in a rough sense. Consistency of the
hadronic picture for equilibration requires that multi-hadron
processes changing the numbers of $\Omega,\Sigma$ etc. must be fast
enough in order to build up the observed particle numbers at $T_{ch}$.
(We do not assume in this part thermal equilibrium with detailed
balance of individual rates.) Keeping in mind the considerable
quantitative uncertainties we now proceed to evaluate rates for
scattering processes involving more than two incoming particles and
demonstrate the importance of such processes near T$_c$.  For an
understanding of multi-particle interactions we write for the rate of
scattering events per volume with $n_{in}$ ingoing and $n_{out}$
outgoing particles
\begin{equation}\label{A}
r(n_{in},n_{out}) = \bar{n}({\rm{T}})^{n_{in}}|{\cal{M}}|^2\phi\\
\end{equation}
with
\begin{equation}\label{AA}
\phi = \prod^{n_{out}}_{k=1}\left(\int\frac{d^3p_k}{(2\pi)^3(2E_k)}
\right)(2\pi)^4\delta^4
\left(\sum_kp^{\mu}_k\right).
\end{equation}
The rate is proportional to $n_{in}$ powers of the particle densities
of the incoming particles that we denote for short by
$\bar{n}(\rm{T})$.  For the outgoing particles $\phi$ is the Lorentz
invariant phase space factor which we evaluate case-by-case with the
program given by \cite{kajantie}, and which needs to be weighted (see
below) with the thermal probability to find a particular cm energy in
the initial state.  The magnitude of the squared transition
amplitude\footnote{We use a normalization adapted to incoming
fermions and outgoing bosons.  For each outgoing fermion the
additional factor $2E$ in the phase space integrals is absorbed here
in the squared amplitude. For each incoming boson there is an
additional factor $2M$ in $|{\cal M}|^2$ which is essentially
cancelled by an additional factor $1/(2 E_i)$ in $r$.} $|{\cal{M}}^2|$
is evaluated using measured cross sections.  We assume constant
$|{\cal{M}}^2|$ independent of temperature and density.

Inspection of measured cross section systematics shows that production
cross sections of strange particles are at most a few mb, and usually
much smaller, as is reflected in the known strangeness suppression
factor in hadronic collisions. Strangeness exchange reactions may
reach cross sections in the 20 mb range.  As cases in point we
evaluate, using eq.~(\ref{A}), in the following the rate $r_{\Omega}$
for $\Omega$ production through the reaction $2\pi+3K \to
\bar{N}\Omega$ and similar rates for $\Xi$ and $\Lambda$ production.

For this case, 
\begin{equation}\label{B}
r_{\Omega} = n_{\pi}^5 (n_{K}/n_{\pi})^3
|{\cal{M}}|^2\phi.
\end{equation}
The densities used to evaluate this expression are taken from the
thermal model predictions which describe the measured yields at RHIC
\cite{rhic} using the excluded volume correction. At T=176 MeV these
are: $n_{\pi}=0.174/$fm$^3$ and $n_K/n_{\pi}$=0.172. Leaving out the
excluded volume correction would increase the densities by
approximately a factor of 2. Note that $n_\pi$ and $n_K$ stand here
for ``generic pion and kaon densities''.  Indeed, the incoming
particles can include all sorts of resonances like $\rho$ etc.
Therefore $n_\pi$ and $n_K$ are not the thermal value of individual
pion and kaon densities but rather comprise effectively all
non-strange and strange degrees of freedom. The $2\pi 3K$ reaction is
likely to be an important channel for $\Omega$ production as it
involves particles with the largest densities (pions, kaons) and the
required amount of strangeness. Similarly we obtain $\Xi$'s from $3\pi
2K$ and $\Lambda$'s from $4\pi K$ reactions.

The numerical rate evaluation needs as input both the matrix element
and phase space factor. The phase factor $\phi$ depends on $\sqrt{s}$
and needs to be weighted by the probability $f(s)$ that the
five-meson-scattering occurs at a given value of $\sqrt{s}$. For this
purpose we assume thermal momentum distributions for the kaons and
pions in the entrance channel. The function $f$ is evaluated
numerically by a Monte-Carlo program. Its results were cross-checked
for massless particles against an analytic evaluation
\cite{knoll}. The phase space factor is then obtained by folding
$\phi$ with $f$ in the energy range from threshold to infinity.

For an estimate of the matrix element we note that the cross section
for $p+\bar{p} \to 5\pi$ has been measured. Close to threshold
it takes a value of about 40 mb and is falling exponentially with cm
energy $\sqrt{s}$ according to $\sigma_{5\pi}=871 {\rm mb}\cdot
\exp{(-\sqrt{s} ~1.95/\rm{GeV})}$ \cite{raf1}. For the $\Omega +
\bar{N}\to 2\pi + 3K$ reaction the threshold is 2.61 GeV and
the peak of $f(s)$ occurs at 3.25 GeV. To evaluate the corresponding
matrix element we assume that the cross section at $\sqrt{s}=3.25$ GeV
is 6.4 mb, close to the $p\bar{p}\to 5\pi$ cross section at
the same energy above threshold. From the known cross section and
phase space the matrix element $|{\cal M}|^2$ can be extracted by the
usual formula. Using our normalization convention this yields
$|{\cal{M}}|^2=9.5\cdot 10^9$/GeV$^8$.  The final result for $\Omega$
production through this channel is then $r_{\Omega}=1.39 \cdot
10^{-4}$/(fm$^4$) at T = 176 MeV and scales with the 5th power of the
pion density. Furthermore, $f$ scales approximately \cite{knoll} as
$\sqrt{s}^{23/2} \exp{-\sqrt{s}/\rm{T}}$, leading to a further
increase of $r_{\Omega}$ with temperature (and hence density).

We have alternatively evaluated the rate for $\Omega$ production in a
semi-classical approach, in which the standard two-body rate equation
$r =n_1 n_2 \langle \sigma v \rangle$ is generalized to multi-particle
collisions. Inspired by the approach taken in cascade codes a reaction
takes place if $n_{in}$ particles approach within a volume V =$4\pi/3
\sqrt{(\sigma_{inel}/\pi)}^3$. For the inelastic cross section we take
a typical value of $\sigma_{inel}=$ 40 mb. A particular exit channel
(such as $\Omega \bar{N}$) is obtained by multiplying with a
probability P$_x$. For the reaction under consideration this yields

\begin{eqnarray}\label{C}
r_{\Omega}&=& 8 P_xn_{\pi}^5 (n_{K}/n_{\pi})^3
\sqrt{(\sigma_{in}^{11}/\pi^3)} \langle v \rangle.
\end{eqnarray}
Using for typical relative velocities $\langle v \rangle = 0.6$ and
P$_x$= 0.10 yields $r_\Omega = 1.4 \cdot 10^{-4}$/fm$^4$, indicating
that a similar result as above is obtained with reasonable parameters.

The meaning of this result is as follows: for a density
$n_{\pi}=0.174$/fm$^3$, as used above, the final $\Omega$ density of
$3.0 \cdot 10^{-4}$/fm$^3$ can be built up within a characteristic
time $\tau_\Omega=n_\Omega/r_\Omega=2.2$~fm. Since the $\Omega$ yield
scales in this approach as $n_{\pi}^5$, and taking into account the
temperature dependence of $f$, already an increase of $n_{\pi}$ to
0.2/fm$^3$ reduces this time by more than a factor of three. The time
$\tau_\Omega$ is depicted in Fig.~\ref{tau} as function of the pion
density. We have also added a temperature scale in this figure. This
scale is obtained as follows: we take the variation of energy density
with temperature from the lattice results (see Fig.~\ref{lattice}) and
assume that the density scales\footnote{Part of the increase in
$\epsilon$ is due to the increase in the number of effective degrees
of freedom. Similarly, more relevant channels for five particle
scattering open up. As argued above $n_\pi$ should be interpreted as
an average weighted density $\bar{n}$ of all particles contributing to
five particle scattering with $\Omega$-production.} as
$\epsilon$/T. Fixing the overall scale by $n_{\pi}=0.174/$fm$^3$ at T
= 176 MeV as above this determines the T-dependence of $n_{\pi}$ near
T$_c$. In consequence, close to T$_c$, the $\Omega$ equilibration time
scales approximately as $\tau_{\Omega} \propto \rm{T}^{-60}$.

We note that the above time of 2.2 fm is a reasonable time for
hadronization and phase change, considering the necessary decrease by
about a factor of 3 in the degrees of freedom and the concomitant
volume increase.  A decrease of the pion density by 1/3 only will
increase this time to about 27 fm; this time is even much longer than
the total lifetime of the fireball as measured by two-pion
correlations \cite{ceres2pi1} while we should consider here only the
time between begin of hadronization and chemical freeze-out.

We have checked the numerics of our approach by noting that, at
equilibrium, detailed balance can be used to evaluate the $2\pi
3K\to \Omega\bar n$ rate in the reverse direction\footnote{We
thank C. Greiner and I. Shovkowy for pointing this out.}. Since both
sides scale very differently with (overall) density, setting both
rates equal determines the equilibrium density (prior to resonance
decays) for each temperature. The so-determined equilibrium density is
to within 25\% equal to that computed independently with our thermal
model, lending strong support to our calculations.

Our main results are fairly insensitive to the details of the
calculation. As can be seen by inspecting Fig.~\ref{lattice} the
energy (and consequently particle) density is very rapidly increasing
with temperature near T=176 MeV due to the phase transition.  Density
values of 20 \% different from those used in our calculations are
reached through temperature changes of less than 3 MeV. Possible
uncertainties in our rate estimates of even a factor of 2.5 would be
compensated by such an increase or decrease in density, indicating the
stability of our estimate near T$_c$. Furthermore, our rate estimates
above are more likely overestimates because of the use of a
comparatively large $\Omega$ N $\to 3K 2\pi$ cross
section. The resulting larger densities needed for chemical
equilibration are, however, easily reached as in our picture the phase
transition is passed through from above.

We have, exactly along the lines for $\Omega$ production, evaluated
the equilibration times for $\Xi$ and $\Lambda$ production with values
of $\tau_{\Xi}=0.71$fm/c and $\tau_{\Lambda}=0.66$fm/c, indicating
that all strange baryons have similar equilibration times with similar
density and temperature dependencies.The corresponding time for
protons and antiprotons is typically shorter.  We conclude that for
all particle species the multiparticle rates are sufficient to produce
the equilibrium abundances.

It is an interesting question if for some particle species like the
pion-proton-antiproton system the hadronic multiparticle rates could
be sufficient to maintain (restricted) chemical equilibrium even for
some temperature range below T$_c$. In this case the measured chemical
temperature for the proton to antiproton ratio should reflect a lower
T$<$T$_{ch}$, which is not the case observationally. This may be taken
as an experimental indication that an extended period of hadronic
equilibrium for T$<$T$_c$ is not realized. It may favour the idea that
the relevant prethermalization process could be associated to
hadronization.

We note that thermal models have also been used \cite{beca3} to
describe hadron production in e$^+$e$^-$ and hadron-hadron collisions,
leading to temperature parameters close to 170 MeV. Indeed,this
suggests that hadronization itself can be seen as a prethermalization
process. However, to account for the strangeness undersaturation in
such collisions, multi-strange baryons can only be reproduced by
introducing a strangeness suppression factor of about 0.5, leading to
a factor of 8 suppression of $\Omega$ baryons. In contrast to heavy
ion collisions, $\tau_\Omega$ exceeds here the available time. In the
hadronic picture this is due to the ''absence'' of sufficient
multi-particle scattering since the system is not in a high density
phase due to a phase transition.

Has the critical temperature of the QCD-phase transition been fixed by
observation?  The answer to this question needs a quantitative
estimate of the difference $\Delta \rm{T=T}_c-\rm{T}_{ch}$. An
accurate determination is difficult since it involves the detailed
understanding of equilibration/ chemical freeze-out. From
Fig.~\ref{tau} we conclude that a temperature decrease of $\Delta {\rm
T}=5$~MeV below the critical temperature lowers the five meson
contribution to $r_\Omega$ by more than a factor of 10. This factor is
even larger if scattering processes with more than five incoming
mesons dominate at T$_{ch}$. Unless strangeness exchanging rates are
unplausibly high at T$_c$, such a sharp drop should make sufficient
$\Omega$ production impossible and we conclude $\Delta \rm{T}
{\stackrel <\sim} 5$~MeV.

The accuracy of the experimental determination of T$_c$ could be
limited by a possible temperature dependence of the hadron masses.
Indeed, the hadronic yields determine the ratio T$_{ch}/M$ rather than
the absolute value of T$_{ch}$. A large uncertainty in $M(T)$ would
reflect in a large uncertainty in $T_{ch}$ and we want to limit the
size of this effect.  Using chiral symmetry arguments one may surmise
that the masses of hadrons except for pions and kaons are proportional
to the value of the order parameter $\sigma$ responsible for chiral
symmetry breaking. As a result, they depend on T and $\mu$ even in the
hadronic phase \cite{CWCHPT}
\begin{equation}\label{x1}
M_j({\rm{T}})=h_j(\rm{T},\mu)\sigma(\rm{T},\mu)
\end{equation}
Neglecting the T-and $\mu$-dependence of the dimensionless couplings
$h_j$ these masses scale proportional to $\sigma(\rm{T},\mu)$.
However, not all hadron masses scale proportional $\sigma(\rm{T},\mu)$
- the pions and kaons scale differently.  Therefore, too large mass
changes are not easily consistent with the universality of chemical
freeze-out. We notice that the ``observed temperature''
T$_{obs}=176$~MeV is fitted to the vacuum masses. The true freeze-out
temperature T$_{ch}$ therefore obeys
\begin{equation}\label{x2}
\frac{\sigma(\rm{T}_{ch},\mu)}{\rm{T}_{ch}}=\frac{\sigma(0,0)}{\rm{T}_{obs}}.
\end{equation}
Since $\sigma$ is expected to decrease for T$>0$ and $|\mu|>0$ (see
e.g. \cite{CWCHPT}) we infer that the true freeze-out temperature
T$_{ch}\approx \rm{T}_c$ is lower than T$_{obs}$.  Already a moderate
relative change of $\sigma$ by 10 \% lowers T$_{ch}$ by 18~MeV
\cite{zschiesche}. Keeping the rather vague character of our ``error
estimate'' in mind we infer the critical temperature for small $\mu$
\begin{equation}\label{9}
{\rm T}_c=176^{+5}_{-18}~{\rm MeV}.
\end{equation}

A continuous crossover, second order phase transition or extremely
weak first order phase transition make a ratio
$\sigma(\rm{T}_{ch})/\sigma(0)$ close to one rather unlikely. The
observed thermal distribution of the hadron yields strongly suggests,
however, that the temperature dependence of the effective hadron
masses should not be too substantial. This even could be interpreted
as an experimental indication that the QCD-phase transition is of
first order! An accurate determination of the temperature dependence
of hadron masses $M(\rm{T})$$/M(0)$ by lattice simulations could
greatly reduce the (lower) error in our estimate of T$_c$.

In summary, hadronic many body collisions near T$_c$ can concistently
account for chemical equilibration at RHIC energies and lead to
T$_{ch}$=T$_c$ to within an accuracy of a few MeV. Any hadronic
equilibrium scenario with T$_{ch}$ substantially smaller than T$_c$
would require that either multi-particle interactions dominate even
much below T$_c$ or that the two-particle cross sections are bigger
than in the vacuum by a large factor. Both of the latter hypotheses
seem unlikely in view of the rapid density decrease.

It was proposed \cite{stock,heinz2,heinz1} that the observed hadron
abundances arise from a direct production of strange (and non-strange)
particles by hadronization. How this happens microscopically is
unclear.  Nevertheless, in order to escape our argument that
T$_{ch}$=T$_c$ one would have to argue that no (even rough) hadronic
picture for this process exists at all - this is unlikely since the
abundances are determined by hadronic properties (masses) with high
precision. Second, one may question if the ``chemical temperature''
extracted from the abundances is a universal temperature which also
governs the local kinetic aspects and can be associated with the
critical temperature of a phase transition in equilibrium. Indeed, in
a prethermalization process, different equilibrium properties are
realized at different time scales. Nevertheless, all experience shows
that kinetic equilibration occurs before chemical equilibration. It
seems quite hard to imagine that chemical equilibrium abundances are
realized at a time when the rough features of kinematic distributions
(like relation between particle density and average kinetic energy per
degree of freedom) are not yet close to their equilibrium
values. Defining the kinetic temperature by the average kinetic energy
one expects that the chemical and kinetic temperatures coincide at
chemical freeze-out (with a typical precision at the percent
level). More specific properties, like detailed balance of hadronic
processes, may be realized later or never.

The critical temperature determined from RHIC for T$_{ch}\approx T_c$
coincides well with lattice estimates \cite{karsch} for $\mu=0$.  The
same arguments as discussed here for RHIC energy also hold for SPS
energies: it is likely that also there the phase transition drives the
particle densities and ensures chemical equilibration. The values of
T$_{ch}$ and $\mu$ collected in \cite{review} would thus trace out the
phase boundary for these energies. Whether the phase transition also
plays a role a lower beam energies is currently an open question.

We thank J. Knoll for helpful discussions concerning phase space
integrals. PBM and JS acknowledge the hospitality of the INT Seattle, where
part of this work was performed.

\end{document}